\documentclass[sigconf]{acmart}
\pdfoutput=1

\usepackage{balance}
\def\BibTeX{{\rm B\kern-.05em{\sc i\kern-.025em b}\kern-.08emT\kern-.1667em\lower.7ex\hbox{E}\kern-.125emX}}

\usepackage{graphicx}
\usepackage{amsmath}
\usepackage{amsfonts}
\usepackage{color}
\usepackage{subcaption}
\usepackage{booktabs}
\usepackage{balance}

\begin{document}
	
\fancyhead{}	

\title{Multi-Agent Reinforcement Learning for Order-dispatching via\\ Order-Vehicle Distribution Matching}

\author{Ming Zhou$^{1}$, Jarui Jin$^1$, Weinan Zhang$^1$, Zhiwei Qin$^2$, Yan Jiao$^2$,\\ Chenxi Wang$^2$, Guobin Wu$^3$, Yong Yu$^1$, Jieping Ye$^2$}
\affiliation{
	$^1$Shanghai Jiao Tong University, $^2$DiDi AI Labs, $^3$DiDi Research
}
\email{
	{zhouming, yyu}@apex.sjtu.edu.cn,
	jinjiarui97@gmail.com,
	wnzhang@sjtu.edu.cn,
	{qinzhiwei, yanjiao, wangchenxi, wuguobin, yejieping}@didiglobal.com
}

\renewcommand{\shortauthors}{Ming Zhou, et al.}
%\def\acmBooktitle#1{\gdef\@acmBooktitle{#1}}
%\acmBooktitle{Proceedings of \acmConference@name
%       \ifx\acmConference@name\acmConference@shortname\else
%         \ (\acmConference@shortname)\fi}

%
% The abstract is a short summary of the work to be presented in the article.
\begin{abstract}
	Improving the efficiency of dispatching orders to vehicles is a research hotspot in online ride-hailing systems. Most of the existing solutions for order-dispatching are centralized controlling, which require to consider all possible matches between available orders and vehicles. For large-scale ride-sharing platforms, there are thousands of vehicles and orders to be matched at every second which is of very high computational cost. In this paper, we propose a decentralized execution order-dispatching method based on multi-agent reinforcement learning to address the large-scale order-dispatching problem. Different from the previous cooperative multi-agent reinforcement learning algorithms, in our method, all agents work independently with the guidance from an evaluation of the joint policy since there is no need for communication or explicit cooperation between agents. Furthermore, we use KL-divergence optimization at each time step to speed up the learning process and to balance the vehicles (supply) and orders (demand). Experiments on both the explanatory environment and real-world simulator show that the proposed method outperforms the baselines in terms of accumulated driver income (ADI) and Order Response Rate (ORR) in various traffic environments. Besides, with the support of the online platform of Didi Chuxing, we designed a hybrid system to deploy our model.
\end{abstract}

\setcopyright{acmcopyright}
\copyrightyear{2019}
\acmYear{2019}
% adjust this to the correct options per the rightsreview. Provided in ACM rightsreview confirmation email.
\acmConference[CIKM '19] {The 28th ACM International Conference on Information and Knowledge Management}{November 3--7, 2019}{Beijing, China}
% \acmBooktitle{The 28th ACM International Conference on Information and Knowledge Management (CIKM'19), November 3--7, 2019, Beijing, China}
\acmPrice{15.00}
\acmDOI{10.1145/3357384.3357799}
% edit the X's to your assigned DOI. Providing in ACM rightsreview confirmation email.
\acmISBN{978-1-4503-6976-3/19/11}


%\ccsdesc[300]{Theory of computation~Multi-agent learning}

% Keywords. The author(s) should pick words that accurately describe the work being
% presented. Separate the keywords with commas.

%
% A "teaser" image appears between the author and affiliation information and the body 
% of the document, and typically spans the page. 
%  \Description{Enjoying the baseball game from the third-base seats. Ichiro Suzuki preparing to bat.}

%
% This command processes the author and affiliation and title information and builds
% the first part of the formatted document.
\maketitle

\section{Introduction}
    With the booming of mobile internet, it becomes feasible and promising to establish the modern large-scale ride-hailing systems such as Uber, Didi Chuxing and Lyft which allow passengers book routes with smartphones and match available vehicles to them based on intelligent algorithms. To some extent, these ride-hailing systems improve the efficiency of the transportation system.

\begin{figure}
        \centering
        \includegraphics[width=.7\columnwidth]{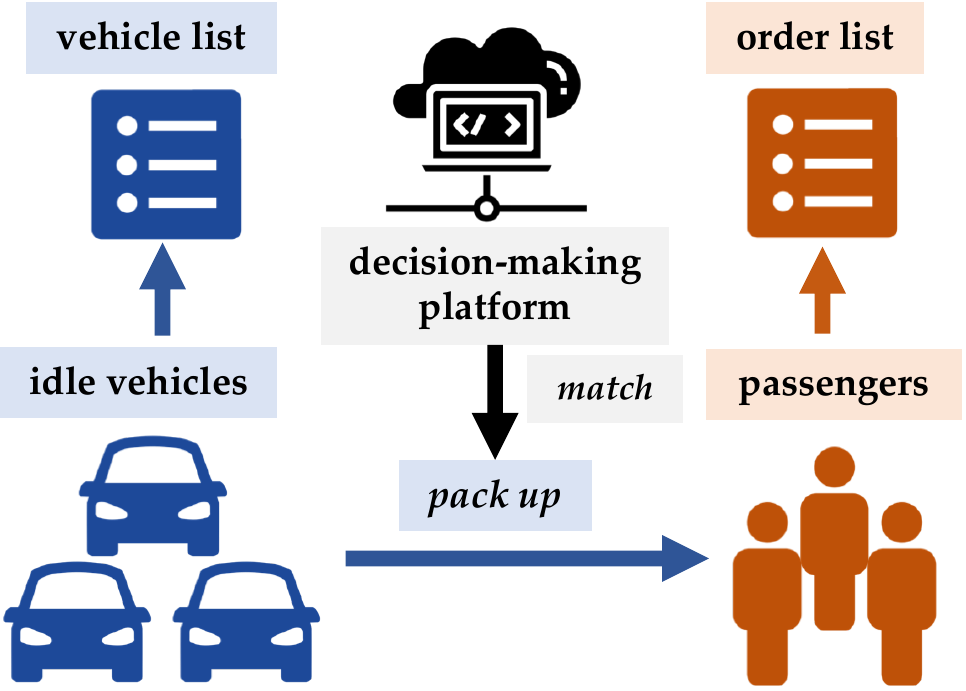}
        \caption{Ride-hailing order-dispatching process}
        \label{fig:order_dispatching}
    \end{figure}

    In ride-hailing systems, a key point is how to dispatch orders to vehicles to make the system work more efficiently and generate more impact. We illustrate the order-dispatching in Figure.~\ref{fig:order_dispatching}, where one can see that the algorithm used by the decision maker is critical for finding suitable matches because the result of order-dispatching has direct influences on the platform efficiency and income. 

    The general strategies of automatically order-dispatching systems are to minimize the waiting time and taxi cruising time through route planning or matching the nearest orders and vehicles \cite{chung2005gps,lee2004taxi,myr2013automatic,chadwick2015context}. 
    In recent research, another approach to solve the order-dispatching problem is to leverage combinatorial optimization \cite{papadimitriou1998combinatorial} to improve the success rate of order-dispatching \cite{zhang2017taxi}. It makes a significant improvement in the online test, but it suffers from high computational cost, and strongly relies on appropriate feature engineering. 
    More importantly, the above strategies are myopic: they may find suitable matches in the current stage, but ignore the potential future impact.

    In this paper, we focus on developing a method to maximize the \emph{accumulated driver income} (ADI), i.e., the impact of orders served in one day, and the \emph{order response rate} (ORR), i.e., the proportion of served orders to the total orders in one day.
    Intuitively, matching vehicles with high-price orders can receive high impact at a single order-dispatching stage. However, if the served orders result in the mismatch between the orders and vehicles in the near future, it would harm the overall service quality in terms of ORR and the long-term ADI.
    Hence, in order to find a balance between the long-term ADI and ORR, it is necessary to develop an order-dispatching algorithm which takes the future supply and demand into consideration.

    \citet{xu2018large} proposed a planning and learning method based on decentralized multi-agent deep reinforcement learning (MARL) and centralized combinatorial optimization to optimize the long-term ADI and ORR. The method formulates the order-dispatching task into a sequential decision-making problem and treats a vehicle as an agent. However, for centralized approaches, a critical issue is the potential "single point of failure" \cite{lynch2009single}, i.e., the failure of the centralized authority control will fail the whole system \cite{lin2018efficient}. Another two related work using multi-agent to learn order-dispatching is based on mean-field MARL \cite{li2019efficient} and knowledge transferring \cite{wang2018deep}.
    
    There are some challenges to be solved when we apply the MARL to the real-time order-dispatching scenario. First, handling the non-stationary environment in MARL is a major problem, which means that all agents learn policies concurrently, while each individual agent does not know the policies of other agents \cite{hu1998multiagent}. The state transition in a multi-agent environment is driven by all agents together, so it is important for agents to have knowledge about other agents' policies. In the order-dispatching scenario, we only care about the idle status of a vehicle since they are available for order-dispatching. However, as the duration of each order is non-deterministic, compared to the traditional multi-agent scenarios which have deterministic time interval, it is difficult to learn the interactions between agents in successive idle states, which makes many MARL methods including opponent modeling \cite{schadd2007opponent,billings1998opponent} and communication mechanism \cite{foerster2016learning} hard to work well.
    Second, the number of idle vehicles keeps changing during the whole episode, i.e., there will always be some vehicles getting offline or online, thus the general MARL methods which require fixed agent number cannot be directly applied in such a case \cite{foerster2016learning2,zheng2017magent}.
    
    In addition, we believe that a higher ORR usually means a higher ADI, and if we can maintain a higher long-term ORR, we will get a higher long-term ADI. With regard to the correctness of this point, we also conducted a corresponding experimental analysis in Section~\ref{sec:lambda}.

    To the best of our knowledge, this is the first work that utilizes this character to improve both ORR and ADI. In detail, we propose a centralized learning and decentralized execution MARL method to solve the above challenges with an extension of Double Q-learning Network \cite{mnih2015human} with Kullback-Leibler (KL) divergence optimization. Besides, the KL-based backward learning optimization method also speeds up the agents learning process with the help of others'. Considering the large scale of agents, and they are homogeneous, we learn only one network using parameter sharing, and share learning experiences among all agents at the training stage, as that in \cite{zheng2017magent,sukhbaatar2016learning}. To address the non-stationary action space problem, in our implementation, the input of deep Q-learning network consists of the state and selected action.

    Extensive experiments with different traffic and order conditions and real-world simulation experiments are conducted. The experimental results demonstrate that our method yields a large improvement on both ADI and ORR compared to the baseline methods in various traffic environments. We also claim the proposed method is highly feasible to be deployed on the existing order-dispatching platform.
    
\section{Related Work}
    \paragraph{\textbf{Taxi-order Dispatching}} There have been several GPS-based order-dispatching systems to enhance the accuracy, communications, and productivity of taxi dispatching \cite{liao2001taxi,liao2003real,myr2013automatic}. These systems do not offer detailed dispatching algorithms, which means these platforms are more like information sharing platforms, helping vehicles choose orders to serve by offering orders information. Other automatic order-dispatching methods \cite{li2011hunting,miao2016taxi} focus on reducing the pick-up distance or waiting time by finding the nearest orders. While these methods usually fail to reach a high success rate on order-dispatching and ignore many potential orders in the waiting list which may be more suitable for vehicles. Zhang et al. \cite{zhang2017taxi} proposed a centralized control dispatching system based on combinatorial optimization. Although it is a simple method, the requirement of computing all available order-vehicle matches can be of much high computational cost in a large-scale taxi-order-dispatching situation. Moreover, it requires appropriate feature engineering. Thus it greatly increases the system implementation difficulty and human efforts of applying the method in a practical situation.

    \paragraph{\textbf{Multi-agent Reinforcement Learning}} Multi-agent reinforcement learning has been applied in domains like collaborative decision support systems. Different from the single agent reinforcement learning (RL), multi-agent RL needs the agents to learn to cooperate with others. It is generally impossible to know other policies since the learning process of all agents is simultaneous. Thus for each agent, the environment is non-stationary \cite{busoniu2006multi}. It is problematical that directly apply the independent reinforcement learning methods into the multi-agent environment. There are several approaches proposed to relieve or address this problem, including sharing the policy parameters \cite{gupta2017cooperative}, training the Q-function with other agents' policy parameters \cite{tesauro2004extending}, centralized training \cite{lowe2017multi} and opponent modeling \cite{schadd2007opponent,billings1998opponent}. Besides, there are also some methods which use explicit communication to offer a relatively stationary environment for peer agents \cite{sukhbaatar2016learning,foerster2016learning,hausknecht2016cooperation}. In the large-scale multi-agent systems, the non-stationary problem will be amplified. To address this problem, Yang et al. \cite{yang2018mean} proposed a novel method which converts multi-agent learning into a two-player stochastic game \cite{shapley1953stochastic} by applying mean field theory in multi-agent reinforcement learning to make it possible in large-scale scenarios. Since the mean-field MARL method only takes a mean field on states/actions input into consideration, it ignores the agent interactions. Our proposed method provides another way to enable large-scale multi-agent learning and retain the interactions between agents, which makes agents receive global feedback from the next moments and adjust their strategies in time. Furthermore, our proposed method provides a backward stationary learning method and has a rapid reaction to the feedback from the environment.

    \paragraph{\textbf{Multi-agent Taxi Dispatching}} A lot of previous work models the taxi dispatching into multi-agent learning, like \cite{alshamsi2009multiagent}, it divides the city into many dispatching areas, and regards an area as an agent, then uses self-organization techniques to decrease the total waiting time and increase the taxi utilization. NTuCab \cite{seow2010collaborative} is a collaborative multi-agent taxi dispatching system which attempts to increase custom satisfaction more globally, and it can dispatch multiple orders to taxis in the same geographical regions. NTuCab thinks that it is not feasible to compute the shortest-time path for each of a possibly large number of available taxis nearby a customer location since it is computationally costly. We follow these settings in our proposed model and divide the city into many dispatching regions. Each dispatching region is controlled in a given distance, which indirectly limits the maximum waiting time. The NTuCab achieves a significant improvement in reducing the wait time and taxi cruising time, but it is also a computational cost method. Xu et al. \cite{xu2018large} proposed a learning and planning method based on MARL and combinatorial optimization recently, and some other methods \cite{wei2018look,oda2018distributed,lin2018efficient} focus on fleet management to improve the ADI or decrease the waiting time. But considering the current operational ride-sharing scenarios, it is hard to perform fleet management for it is impossible to force drivers to designated regions. The mentioned MARL method\cite{xu2018large} is an independent MARL method, which ignores the interactions between agents. However, it is a consensus to consider that the agent interactions have a positive impact on making optimal decisions. Our proposed method considers the interaction between agents by applying constraints on the joint policies using KL-divergence optimization, and the experiments demonstrate that the proposed method outperforms baselines on all metrics in different traffic environments.

\section{Methodology}\label{sec:method}
%    In this section, we firstly give the problem definition of regarding order-dispatching as a multi-agent reinforcement learning process, then discuss the solutions to solve the main challenges when using the MARL method and give the discussion of our method.
    In this section, we first give a definition of order-dispatching from a perspective of multi-agent reinforcement learning process, and then discuss the main challenges when applying the MARL method for order-dispatching, and give our methods.

\subsection{Order-dispatching as a Markov Game}
    We regard the order-dispatching task as a sequential decision task, where the goal is to maximize the long-term ADI and ORR per day. According to the characters of the practical environment, each vehicle can only serve the surrounding orders, thus we model the order-dispatching task using \textit{Partially Observable Markov Decision Process} (POMDP) \cite{spaan2012partially} in multi-agent settings. With the multi-agent settings, we can decompose the original global order-dispatching task into many local order-dispatching tasks, and transform a high-dimensional problem into multiple low-dimensional problems.

    The POMDP framework to the multi-agent order-dispatching problem can be formulated as a tuple $\langle \mathcal{S},  \mathcal{P}, \mathcal{A}, \mathcal{R}, \mathcal{G}, \mathcal{N}, \gamma \rangle$, where $\mathcal{S}$, $\mathcal{P}$, $\mathcal{A}$, $\mathcal{R}$, $\mathcal{G}$, $\mathcal{N}$, $\gamma$ represent the sets of states, state transition probability function, sets of action spaces, reward functions, set of grids, the number of agents and the future reward discount factor respectively.
    
    For each agent $i$, $\mathcal{S}_i \in \mathcal{S}$, $\mathcal{A}_i \in \mathcal{A}$, $\mathcal{R}_i \in \mathcal{R}$ represent the state space, action space and reward function respectively, and $\mathcal{G}_i \in \mathcal{G}$ represents the grid which the agent in. The state transition occurs after the decision making, i.e. agents executed their actions, then the state $\mathcal{S}_t$ of environment at time $t$ transform to $\mathcal{S}_{t+1}$ at time $t+1$, and agents will get rewards given by the environment. Based on the above definitions, the main purpose of each agent is to learn to maximize the cumulative reward $G_{t:T}$ from $t$ to $T$
    \begin{displaymath}
        \max G_{t:T} = \max \sum_{t=0}^{T}{\gamma^t r_t(s_t,a_t)}~,\mbox{ where } a_t \sim \pi_{\theta}(s_t)~.
    \end{displaymath}

    In reinforcement learning, the $\pi_{\theta}(\cdot)$ parameterized with $\theta$ represents the policy with respect to the state at time $t$. 
    
    It is common to divide the city into regional dispatch areas \cite{lin2018efficient,seow2010collaborative}. In our settings, we use a grid-world to represent the real world and divide the real world into several order-dispatching regions. Each grid represents an individual order-dispatching region which contains some orders and vehicles, and we regard vehicles as agents here. Based on the above MARL settings, we specify the definitions of the order-dispatching task as follows from a mathematical perspective.

    \begin{itemize}
    	\item \textbf{State}: The state input used in our method is expressed as a four elements tuple, namely, $\mathcal{S} = \langle G, N, M, \mathcal{D}_{\text{dest}} \rangle$. Elements in the tuple represent the grid index, the number of idle vehicles, the number of valid orders and the distribution of orders' destinations respectively. The distribution of order's destination is a mean over the destination vectors of orders in grid $G$, which roughly reflects the overall the orders information. In our settings, agents in the same grid share the same state.
    	
    	\item \textbf{Action}: The action input used in our method is expressed as $\mathcal{A} = \langle G_{\text{source}}, G_{\text{dest}}, T, C \rangle$. Elements in the tuple represent the source grid index, target grid index, order duration, and price respectively. We regard the set of orders in the grid $j$ at time $t$ as the candidate actions of the agent $i$. Since agents are homogeneous, so agents in grid $j$ share the same action space.
    	In practice, sometimes there is no order in some regions. Under the setting of MARL, agents need to select orders at each timestep, but some grids may not have orders, so in order to ensure the feasibility and sustainability of the MDP, we artificially add some virtual orders whose $G_{\text{source}}=G_{\text{dest}}$, and set the price $C$ to 0. When idle vehicles select these virtual orders, it means they will stay where they are.
    	
    	\item \textbf{State Transition}: The agent which serves one order will migrate to the destination grid given by the taken order after $T$ time step, where $T$ is defined with the served order duration, then the state of agent will be updated to the newest state, namely, the stage of destination grid.

    	\item \textbf{Reward}: The reward function is very important for reinforcement learning to a great extent which determines the direction of optimization. Because of the goal of learning is to find a solution which maximizes the ADI with high ORR, so we design a reward function which is proportional to the price of each order.
    \end{itemize}

\subsection{Non-stationary Action Space}
    Traditional deep Q-learning network accepts a state input and outputs a vector of Q values whose dimension is equal to the dimension of action space, i.e.,
    \begin{align}
    	\text{dim}\Big(Q(s, \mathcal{A})\Big) = \text{dim}\Big(\mathcal{A}\Big)~.
    \end{align}

    It is correct when the action space is fixed, while it is problematic in our settings. There is a fact that for the grid $j$, the orders produced at time $t$ are always different from the orders produced at other moments. It cannot ensure that the action space is consistent along with the whole episode, so it is problematical to regard the orders as an action while ignoring the distribution of the variant action space. In our proposed method, we use the tuple $\langle S, a \rangle$ to represent the input of Q-learning, then evaluate all available state-order pairs.

\subsection{Action Selection Q-learning}\label{subsec:qlearning}
\begin{figure}[ht]
    	\centering
    	\includegraphics[width=.90\columnwidth]{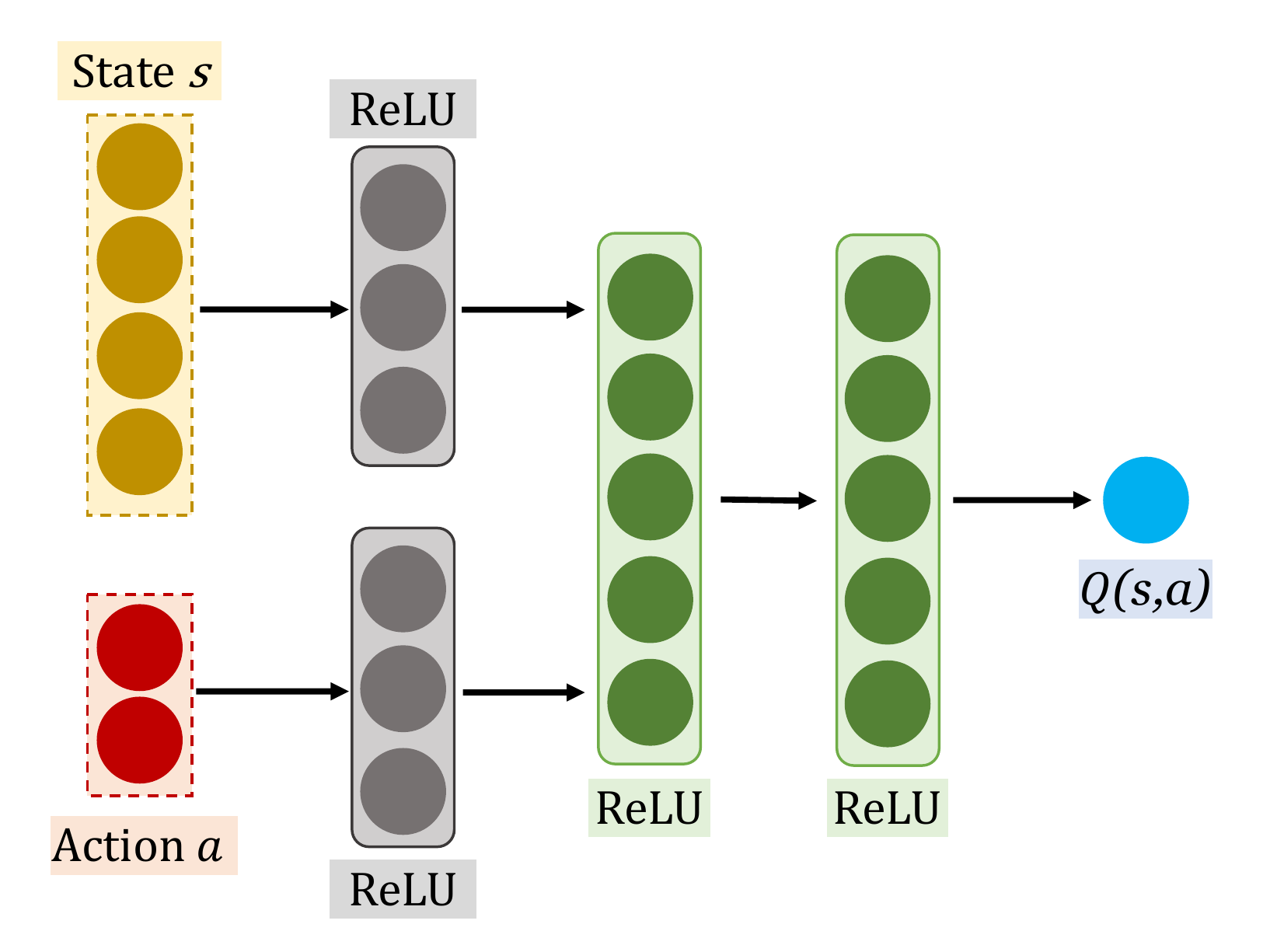}
    	\caption{Action selection Q-learning. Different from the traditional Q-learning network, our model accepts both state input and action feature vector. After embedding them respectively, there is a concatenation followed by 2 dense layers which follows the embedding layers. Then the network outputs a scalar value $Q(s,a)$.}
    	\label{fig:model}
    \end{figure}

    For convenience, we name the Q-learning network with a state-action input as \emph{action selection Q-learning} shown in Figure~\ref{fig:model}.
    
    For agent $i$, supposing there are $M$ available orders, which requires $M$ state-action evaluation. In the case of $N$ agents, the computational complexity will be $\mathcal{O}(N \cdot M)$. To decrease the original complexity to $\mathcal{O}(M)$, we use parameter sharing and state sharing mentioned in previous sections to achieve it.

    From the perspective of agent $i$, we suppose that $s_t$ denotes the state at time $t$, $a_t$ denotes the set of orders, then the Bellman Equation in our settings can be expressed as
    \begin{align}
    	&Q(s_t,a_t) = \alpha Q(s_t,a_t) + \notag\\
    	&\quad(1 - \alpha)\Big[r_{t} + \gamma \cdot \mathbb{E}_{a_{t+1} \sim \pi(s_{t+1})}[Q(s_{t+1}, a_{t+1})]\Big]~,
    	\label{eq:bellman}
    \end{align}
    where $\gamma \in [0,1]$ is the discount factor, $\alpha$ is the step size. The value of the next timestep is a expectation of all available state-order pairs. When the policy $\pi(s_{t+1})$ is greedy, then Eq.~\eqref{eq:bellman} represents the traditional Q-learning algorithm.
%    \begin{align}
%    	&\mathbb{E}_{a_{t+1} \sim \pi(s_{t+1})}[Q(s_{t+1}, a_{t+1})] = \notag\\ &\sum_{\pi(a_{t+1} \mid s_{t+1})} \pi(a_{t+1} \mid s_{t+1}) Q(s_{t+1}, a_{t+1})~.
%    \end{align}

    To balance the exploitation and exploration, the Q values related to the same orders set are converted into a biased strategy \textit{Boltzman exploration}
    \begin{align}
    	\pi(a^j_t \mid s_t) = \frac{e^{Q(s_t,a^j_t) / \tau}}{\sum_{a^j_t \in \mathcal{A}_i}{e^{Q(s_t, a^j_t) / \tau}}}~,
    \end{align}
    where $\tau$ is the temperature to balance the exploitation and exploration.

    \begin{figure}[ht]
    	\centering
    	\includegraphics[width=.8\columnwidth]{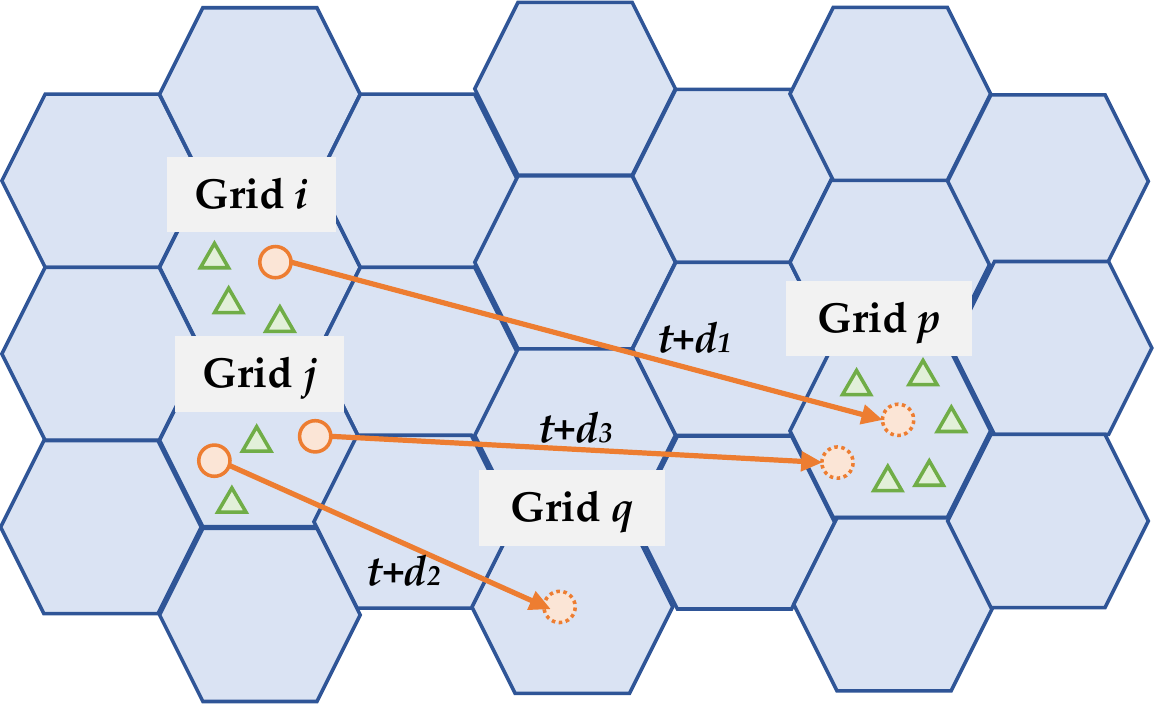}
    	\caption{Grid-based order-dispatching. Supposing that triangles in each grid represent orders, and dots represent vehicles. It shows that the order-dispatching process of each grid at time $t$, and different order has different duration of $d$, so the vehicles will arrive at the destination grids at different time, and vehicles serve different orders will be assigned to different grids, then it is hard to form continuous interactions and communication between vehicles. For these two reasons, applying the communication mechanism or learning others' policies is not a good choice.}
    	\label{fig:hexagonal}
    \end{figure}

\subsection{KL Divergence Optimization}
    In the multi-agent system, the main method to relieve or overcome the non-stationary problem is learning multi-agent communication \cite{hausknecht2016cooperation,foerster2016learning,foerster2016learning2,sukhbaatar2016learning}, while most of them require a fixed agent number or observations from other agents before making decisions. 
    In the order-dispatching case, explicit communication occurs between agents is often time-consuming and difficult to adapt. As illustrated in Figure~\ref{fig:hexagonal}, supposing that triangles in each grid represent orders, and dots represent vehicles. It shows that the order-dispatching process of each grid at time $t$, and different order has different duration of $d$, so the vehicles will arrive at the destination grids at different time, and vehicles serve different orders will be assigned to different grids, then it is hard to form continuous interactions and communication between vehicles. Also, it often suffers from computational cost, especially in large-scale settings.
    Taking the aforementioned reasons, we introduce a centralized training method using KL divergence optimization, which aims to optimize the agents' joint policy and try to match the distribution of vehicles with the distribution of orders.

    Notice that we have two goals need to achieve in our proposed method: \textit{(1) maximize the long horizontal ADI; (2) optimize the order response rate}. If there are always enough vehicles in the dispatching grid, it is easy to decrease the rate of idle vehicles and improve the order response rate, also the long horizontal ADI, while there is a fact that we cannot control the distribution of orders. So we want to make the order and vehicle distribution as similar as possible through finding feasible order-vehicle matches. We do not require explicit cooperation or communication between agents, but an independent learning process with centralized KL divergence optimization.

    Supposing at time $t$, the agents find a feasible order set $\mathcal{O}_t$ by executing their policies, namely,
    \begin{align}
    	\mathcal{O}_t \sim \{\pi_{\theta,j}(s_t) \mid j=1,\dots,N\}~.
    \end{align}

    Our purpose is to find an optimal order set $\mathcal{O^*}_t$. Focusing on a certain grid $j$, it supposes that the policy $\pi_{\theta_j}$ at time $t$ is parameterized by $\theta_j$. After all policies have been executed, we get the newest distribution of vehicles $\mathcal{D}_{t+1}^v$, and the newest distribution of orders is $\mathcal{D}_{t+1}^o$. The KL divergence from $\mathcal{D}_{t+1}^v$ to $\mathcal{D}_{t+1}^o$ shows the margin between the joint policy $\Pi$ at time $t$ and $\mathcal{D}_{t+1}^o$, so the KL optimization is actually finding an optimal joint policy $\Pi^*$ which has a minimal margin:
    \begin{equation}
    	\Pi^* = \arg_{\Pi} \min D_{\textit{KL}}(\mathcal{D}_{t+1}^o \parallel \mathcal{D}_{t+1}^v(\Pi))~,
    \end{equation}
    where $\Pi = \{\pi_{\theta,j} \mid j=1,...,N\}$. For the convenience, we replace $D_{\textit{KL}}(\mathcal{D}_{t+1}^o \parallel \mathcal{D}_{t+1}^v(\Pi))$ with $D_{KL}$. We want to decrease the KL divergence from the distribution of vehicles to the distribution of orders to balance the demand and supply at each order-dispatching grid. Formally, our KL policy optimization can be written as:
    \begin{align}
        &\min_{\theta} \mathcal{L} = \parallel Q_{\theta}(s,a) - Q^* \parallel_2 \\
        &\begin{array}{r@{\quad}r@{}l@{\quad}l}
        s.t.& D_{KL} \le \beta
        \end{array}~,
    \end{align}
    where $\beta \in \mathbb{R}$. Then the objective function can be expressed as
    \begin{align}
    	\min_{\theta} \mathcal{L} = \parallel Q_{\theta}(s,a) - Q^* \parallel_2 +\lambda D_{KL}~,
    \end{align}
    where $Q^*$ is the target Q-value, $\lambda \in \mathbb{R}$ parameterizes the contribution of KL item. To formulate the relationship between $\min \mathcal{L}$ and $\theta_j$, we make some definitions of notations in advance. Considering that there is $N$ grids in total, $n^i_{t+1}$ represents the number of idle vehicles in grid $i$ at time step $t+1$, which can be formulated as $n^i_{t+1} = \sum^N_{j=1}c^j_{t} \cdot \pi_{j \rightarrow i} $, where $c^j_{t}$ represents the idle driver number at last time step $t$, $\pi_{j \rightarrow i}$ represents the probability of dispatching orders which from grid $j$ to grid $i$ to idle vehicles at time $t$, and these vehicles will arrive at grid $i$ at time $t+1$. $q^j_{t+1}$ is the rate of idle vehicles in grid $j$ which can be formulated into $q^j_{t+1} = n^j_{t+1} / \sum_{k=1}^N{n^k_{t+1}}$. $p^i_{t+1}$ represents the rate of orders in grid $i$ at time $t+1$ here. Using chain rule, we can decompose the gradient of $D_{\textit{KL}}$ to $\theta$ as
    {\small
    	\begin{align}
    		&\nabla_{\theta_j} D_{\textit{KL}} = \nabla_{\pi_{j}} D_{KL} \cdot \nabla_{\theta_j} \pi_j \notag \\
    		&= -\Big(\sum^N_{i=1} p^i_{t+1} \nabla_{\pi_j} \log \frac{q^i_{t+1}}{p^i_{t+1}} \Big) \cdot \nabla_{\theta_i} \pi_j \notag \\
    		&= \sum^N_{i=1}p^i_{t+1}\nabla_{\pi_j}\log \frac{1}{q^i_{t+1}} \cdot \nabla_{\theta_j}\pi_j \notag \\
    		&= \sum^N_{i=1} p^i_{t+1} \Big[\nabla_{\pi_j}\log {\sum^N_{k=1}\sum^N_{l=1}\pi_{l \rightarrow k} c^l_t} -\nabla_{\pi_j} \log{\sum^N_{k=1} \pi_{j \rightarrow k} c^j_{t}} \Big] \cdot \nabla_{\theta_j} \pi_j \notag \\
    		&= \sum^N_{i=1} p_i \Big[ \frac{\nabla_{\pi_j}\sum^N_{k=1}\pi_{j \rightarrow k}(c^j_t + \sum^N_{l\ne j}c^l_t)}{\sum^N_{k=1}\sum^N_{l=1}\pi_{l \rightarrow k} c^l_t} - \frac{\nabla_{\pi_j}\sum^N_{k=1}\pi_{j \rightarrow k}c^j_t}{\sum^N_{k=1}\pi_{j \rightarrow k} c^j_t} \Big] \cdot \nabla_{\theta_j} \pi_j \notag \\
    		&= c^j_t \sum^N_{i=1} p^i_{t+1} \Big[ \frac{1}{\sum^N_{k=1}\sum^N_{l=1}\pi_{l \rightarrow k} c^l_t} - \frac{1}{\sum^N_{k=1}\pi_{j \rightarrow k} c^j_t} \Big] \cdot \nabla_{\theta_j} \pi_j \notag \\
    		&= c^j_t\sum^N_{i=1}p^i_{t+1}\Big[ \frac{1}{N_{\textit{vehicle}}} - \frac{1}{n^i_{t+1}} \Big] \cdot \nabla_{\theta_j}\pi_j~,
    	\end{align}
    }
    where $N_{\textit{vehicle}}=\sum^N_{j=1}{n^j_{t+1}}$. The gradient of $\pi_j$ to $\theta_j$ is $\nabla_{Q_j(s,a)}\pi_j(a \mid s) \cdot \nabla_{\theta} Q(s,a)$. We use the $\delta = \parallel Q - Q^* \parallel_2$, then the final gradient of $\mathcal{L}_{\theta}(s,a)$ to $\theta$ is calculated as
    \begin{align}
    	\nabla_{\theta_j} \mathcal{L} = \nabla_{\theta_j}\delta + \lambda \nabla_{\theta_j} D_{\textit{KL}}~.
    \end{align}
    
    For convenience, we give a summary for some important notations in Table~\ref{tab:notation}.
    
    \begin{table}[htbp]\caption{Important notations}
    	\begin{center}% used the environment to augment the vertical space
    		% between the caption and the table
    		\begin{tabular}{r c p{.8\columnwidth} }

    			\toprule
    			$N$ & & the number of grids\\
    			$n^i_{t+1}$ & & the number of idle vehicles in grid $i$ at time $t+1$\\
    			$c^j_t$ & & the number of idle vehicles in grid $j$ at time $t$\\
    			$\pi_{j\rightarrow i}$ & & the probability of dispatching orders which from grid $j$ to grid $i$ to idle vehicle at time $t$\\
    			$q^j_{t+1}$ & & the rate of idle vehicles in grid $j$\\
    			$p^i_{t+1}$ & & the rate of orders in grid $i$ at time $t+1$\\
    			\bottomrule
    		\end{tabular}
    	\end{center}
    	\label{tab:notation}
    \end{table}
    
%    \begin{

\section{Experiments}
    We examine the correctness of our model in a toy grid-based order-dispatching environment and the practicality of our model using real-world data from three cities. Considering the constraint of a grid-based environment, we did not compare with order-dispatching algorithms based on coordinate systems. To compare with existing methods, and investigate the effectiveness of our method on the metrics of ORR and ADI, we select three typical algorithms as baselines, namely, Independent Deep Q-learning Network (IL), Nearest order-dispatching (NOD) and MDP respectively, and we will give a brief description at the first.

    \begin{itemize}
    	\item \textbf{IL}: A variant of Double DQN \cite{van2016deep} which takes a tuple of state and action as an input. Compared with our method, the only difference in IL is that KL optimization is not used.
    	\item \textbf{NOD}: Nearest-distance Order Dispatching (NOD) algorithm, which dispatches orders to idle vehicles with considering the shortest distance. The reason why we use NOD as one of the baselines is that it is a fairly representative algorithm which is used frequently and easy to implement in practice. However, in our environment setting, because there are reasonable regional division strategies (for example, in our later real-world data experiments, the size of the division area guarantees the maximum order waiting time is 10 minute), we have no need to distinguish the specific position of vehicles in the same dispatching region. That is to say, the principle of matching orders based on distance is equivalent to random matching in our environment setting.
    	\item \textbf{MDP}: Proposed by \citet{xu2018large}, a planning and learning method based on decentralized multi-agent deep reinforcement learning and centralized combinatorial optimization.
    \end{itemize}

    Considering the fairness of experiments, we use the same reward function for the reinforcement learning methods.

    \begin{table*}
    	\setlength{\belowcaptionskip}{-0mm}
    	\centering
    	\caption{Performance comparison in terms of ADI and ORR with respect to NOD. We compare against with baselines in three different order distribution changes degree, namely, low, medium and high. KL-Based is our proposed method, which outperforms all baselines on all metrics.}
    	\setlength{\tabcolsep}{1mm}
    	\begin{tabular}{|c|cc|cc|cc|}
    		\hline
    		Order Distribution Divergence & \multicolumn{2}{c|}{Low} & \multicolumn{2}{c|}{Medium} & \multicolumn{2}{c|}{High}\\
    		\hline
    		Metrics & ADI & ORR & ADI & ORR & ADI & ORR\\
    		\hline
    		\hline
    		IL & +12.5\% & +6.94\% & +11.5\% & +6.3\% & +6.68\% & +2.32\% \\
    		MDP & +14.5\% & +8.94\% & +13.3\% & +6.69\% & +7.28\% & +3.42\% \\
    		KL-Based & +\textbf{25.12}\% & +\textbf{13.40}\% & +\textbf{20.94}\% & +\textbf{7.89}\% & +\textbf{13.47}\% & +\textbf{4.61}\% \\
    		\hline
    	\end{tabular}
    	\label{tb:toy_table}
    \end{table*}
    
    \begin{table}
    	\setlength{\belowcaptionskip}{-0mm}
    	\centering
    	\caption{Performance comparison in terms of ADI and ORR with respect to NOD. We compare against baselines using different datasets from three cities. KL-Based is our proposed method, which outperforms all baselines on all metrics.}
    	\setlength{\tabcolsep}{1mm}
    	\begin{tabular}{|c|cc|cc|cc|}
    		\hline
    		City & \multicolumn{2}{c|}{City A} & \multicolumn{2}{c|}{City B} & \multicolumn{2}{c|}{City C}\\
    		\hline
    		Metrics & ADI & ORR & ADI & ORR & ADI & ORR\\
    		\hline
    		\hline
    		IL & +4.69\% & +1.68\% & +2.96\% & +1.11\% & +4.72\% & +2.05\% \\
    		MDP & +5.80\% & +1.89\% & +3.69\% & +2.63\% & +5.98\% & +2.14\% \\
    		KL-Based & +\textbf{6.46}\% & +\textbf{3.07}\% & +\textbf{4.94}\% & +\textbf{3.30}\% & +\textbf{6.12}\% & +\textbf{3.01}\% \\
    		\hline
    		%\bottomrule
    	\end{tabular}
    	\label{tb:real_world_exp}
    \end{table}

\subsection{Model Settings}
    Our model is an extension of Double DQN with soft update. All neural-based models used in our experiments are implemented by the MLP with 2 hidden layers, and the active function used here for all neural-based algorithm is rectified linear unit (ReLU). The replay buffer stores experience tuples, which can be formulated into $\langle s_t, a_t, \mathcal{A}_t, s_{t+1}, \mathcal{A}_{t+1}, \nabla_{\pi} D_{\textit{KL}} \rangle$. Elements of the tuple represent the state of time $t$, action selected at time $t$, the action set at time $t$, state at time $t+1$, action set at time $t+1$ and the first gradient item of $\nabla_{\theta} D_{\textit{KL}}$, respectively. The temperature is 1.0, discount factor $\gamma=0.95$ and learning rate is $\alpha=10^{-4}$.

\subsection{Particle Order-dispatching Experiment}
\begin{figure}[ht]
    	\centering
    	\begin{subfigure}[b]{.48\columnwidth}
    	    \centering
    		\includegraphics[width=\columnwidth]{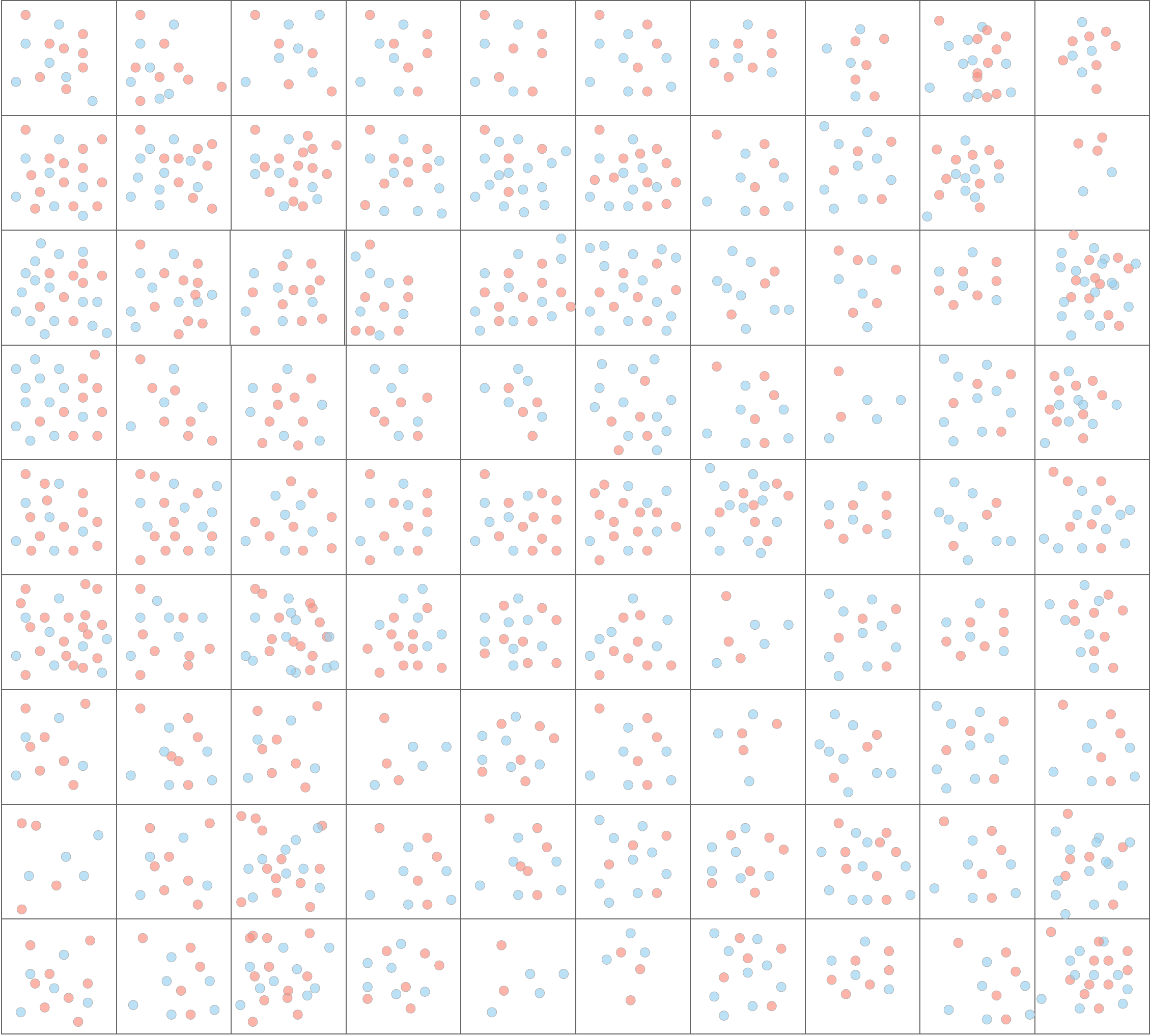}
    		\caption{}
    		\label{subfig:toy_env_big}
    	\end{subfigure}
    	\hfill
    	\begin{subfigure}[b]{.43\columnwidth}
    	    \centering
    		\includegraphics[width=\columnwidth]{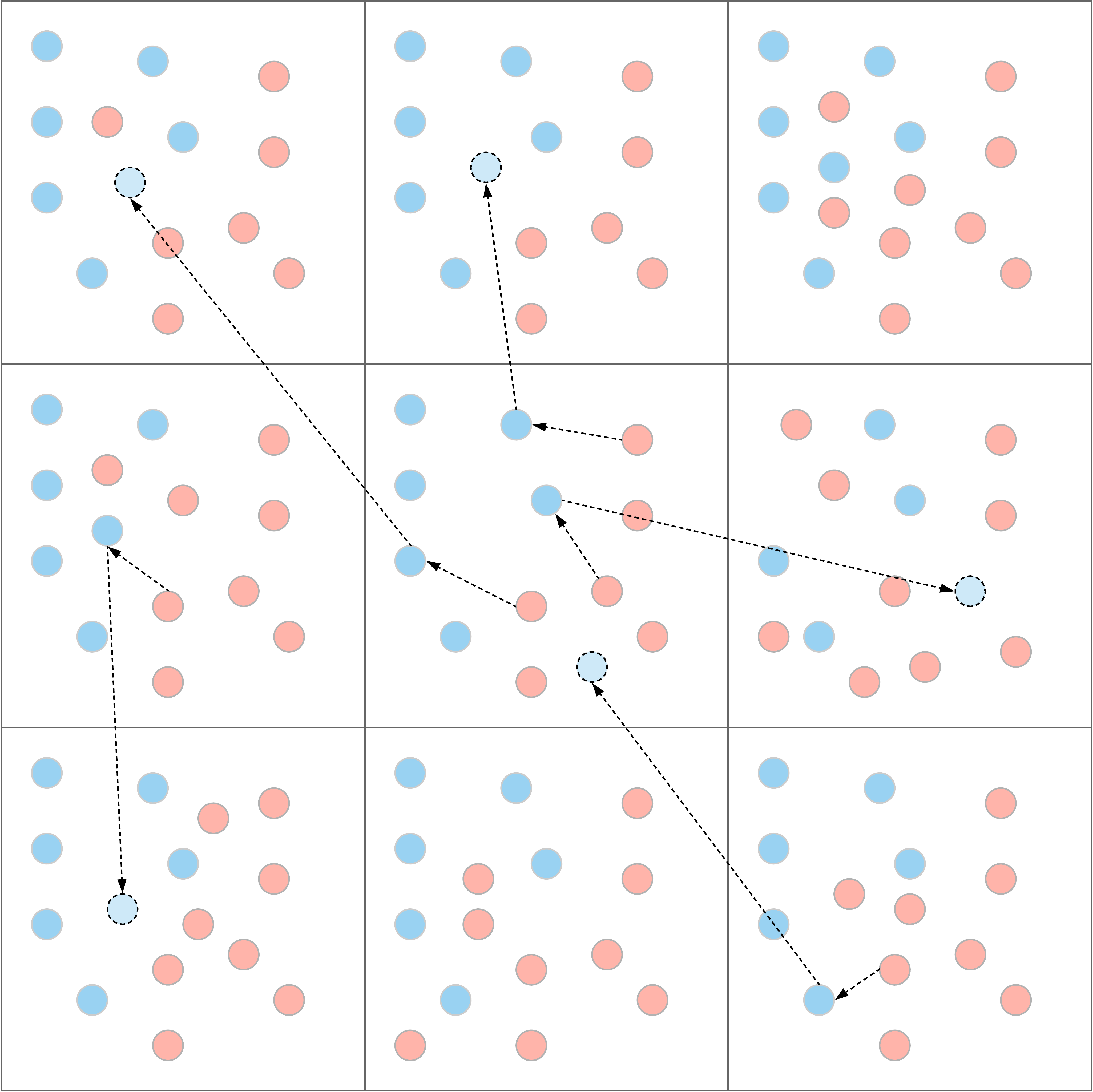}
    		\caption{}
    		\label{fig:toy_env_migrate}
    	\end{subfigure}
    	\caption{The toy order-dispatching environment. (a) The 10x10 toy order-dispatching environment. Red particles and blue particles represent orders and vehicles. (b) The migrate process of blue particles, where the dotted arrow indicates that the red one is dispatched to the blue one, the solid arrow indicates that the blue one will migrate to another grid at the next timestep. For the red particles in each grid, the surrounding 8 grids are feasible destinations.}
    	\label{fig:particle_env}
    \end{figure}

    The grid-based order-dispatching environment showed in Figure~\ref{fig:particle_env} is implemented based on the multi-agent particle environment supported by Mordatch et al. \cite{mordatch2017emergence}. This toy environment abstracts the real-world order-dispatching, where one grid represents one dispatching region, and orders have the same duration. The blue particles and the red particles represent vehicles and orders respectively. All of the blue particles and red particles scatter in a $10 \times 10$ grid world. Each of the red particles owns a direction vector $v$ and a reward $r$, the direction vector $v=\langle \#source, \#target \rangle$ denotes a direction from the source grid to the target grid. In this environment, the order price is simplified with Euler distance between grids, so reward $r$ is proportional to the Euler distance between $\#source$ and $\#target$, i.e.,
    \begin{align}
    	r=0.1 \times \parallel \#source - \#target \parallel_2~.
    \end{align}

    The blue particles will get reward by picking red particles in the same grid, one blue particle can only pick one red particle at each time-step. Blue particles will migrate to the target grids given by the red particles at the next timestep. In our settings, the time horizon is $T=144$. At $t=0$, we produce the reds and blues using specific distributions respectively. At the next timesteps from $t=1$ to $t=T-1$, there are some new red particles born with a specific distribution with respect to the grid position and time, while there is no new blue particles born expect $t=0$, that means all blue particle movements fully dependent on picking red particles. In order to match the real world situation, the amount of blues is less than the reds in our settings.

\subsubsection{Influence of KL Divergence}
\begin{figure}[htb]
        \centering
        \begin{subfigure}[b]{.46\columnwidth}
            \centering
            \includegraphics[width=\columnwidth]{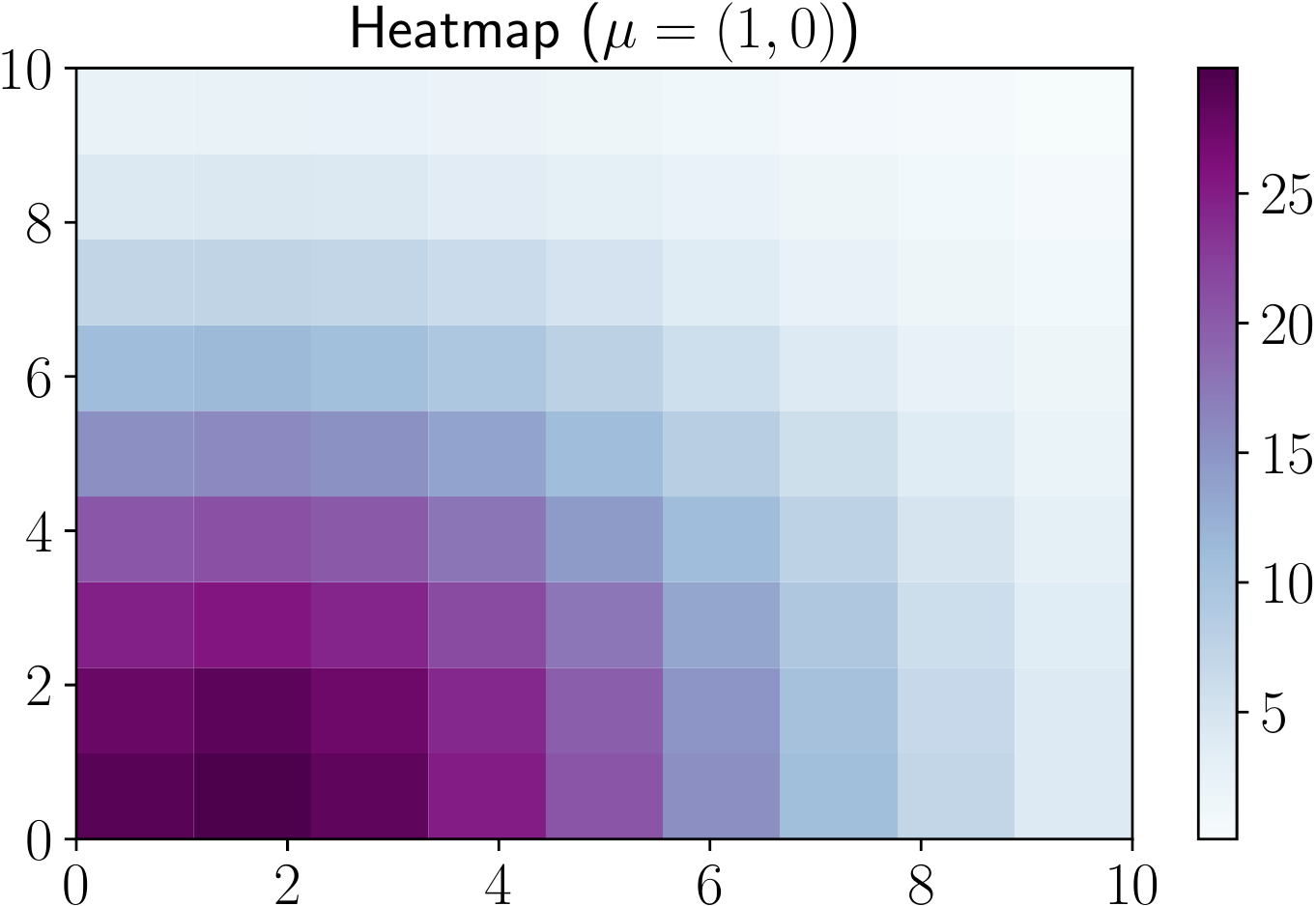}
            \caption{}
            \label{subfig:heatmap1}
        \end{subfigure}
        \hfill
        \begin{subfigure}[b]{.46\columnwidth}
            \centering
            \includegraphics[width=\columnwidth]{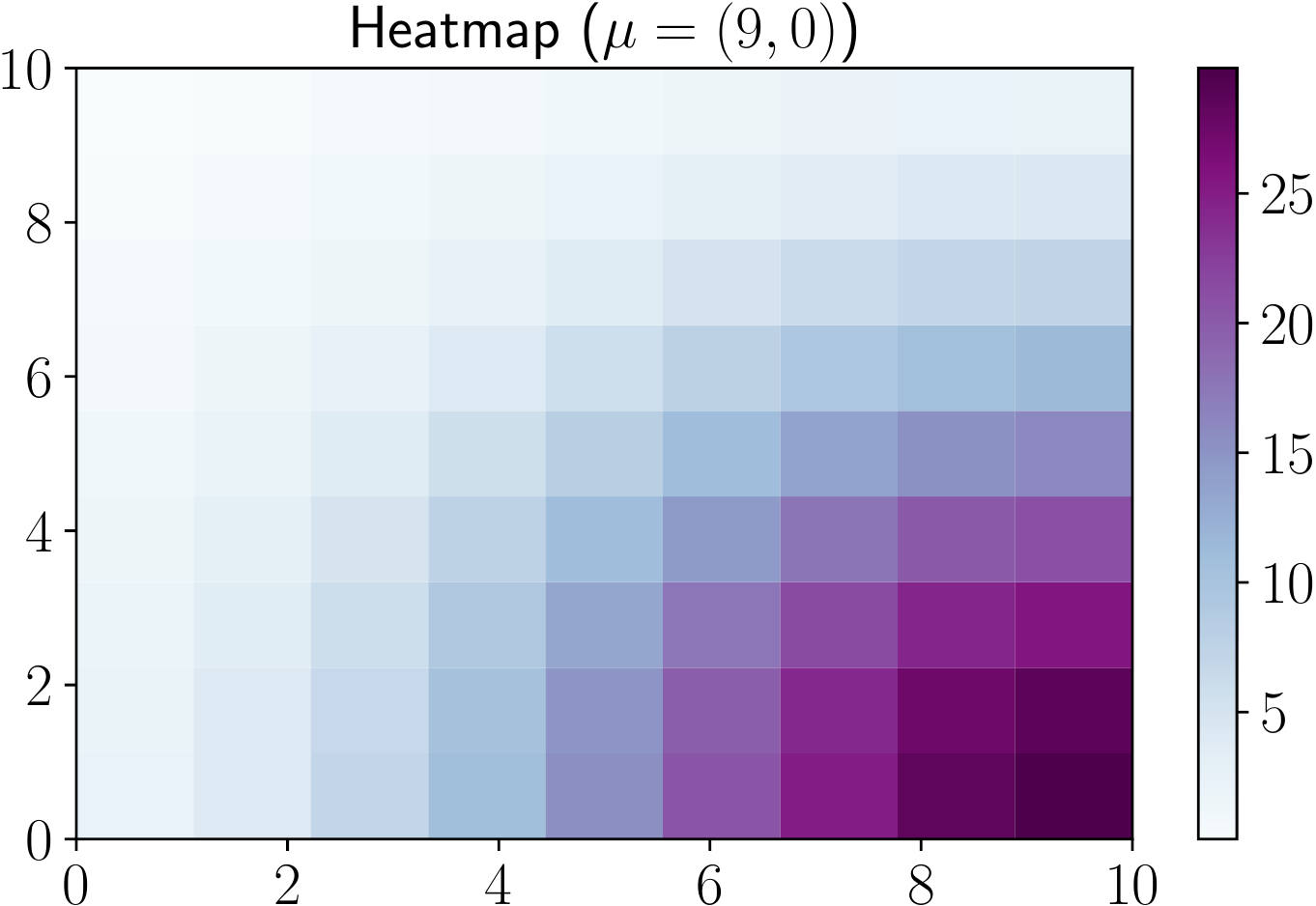}
            \caption{}
            \label{subfig:heatmap2}
        \end{subfigure}
        
        \begin{subfigure}[b]{.46\columnwidth}
            \centering
            \includegraphics[width=\columnwidth]{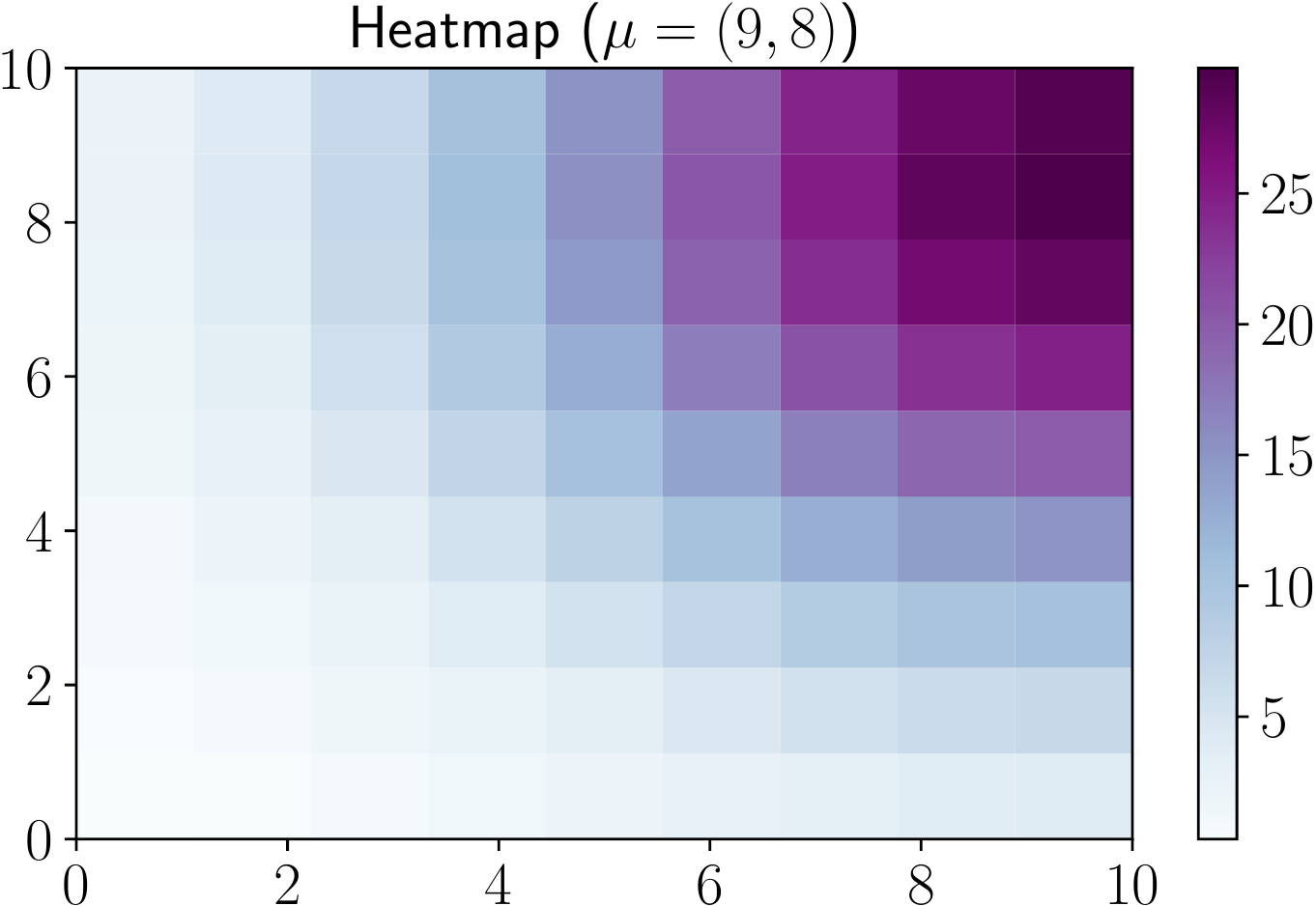}
            \caption{}
            \label{subfig:heatmap3}
        \end{subfigure}
        \hfill
        \begin{subfigure}[b]{.46\columnwidth}
            \centering
            \includegraphics[width=\columnwidth]{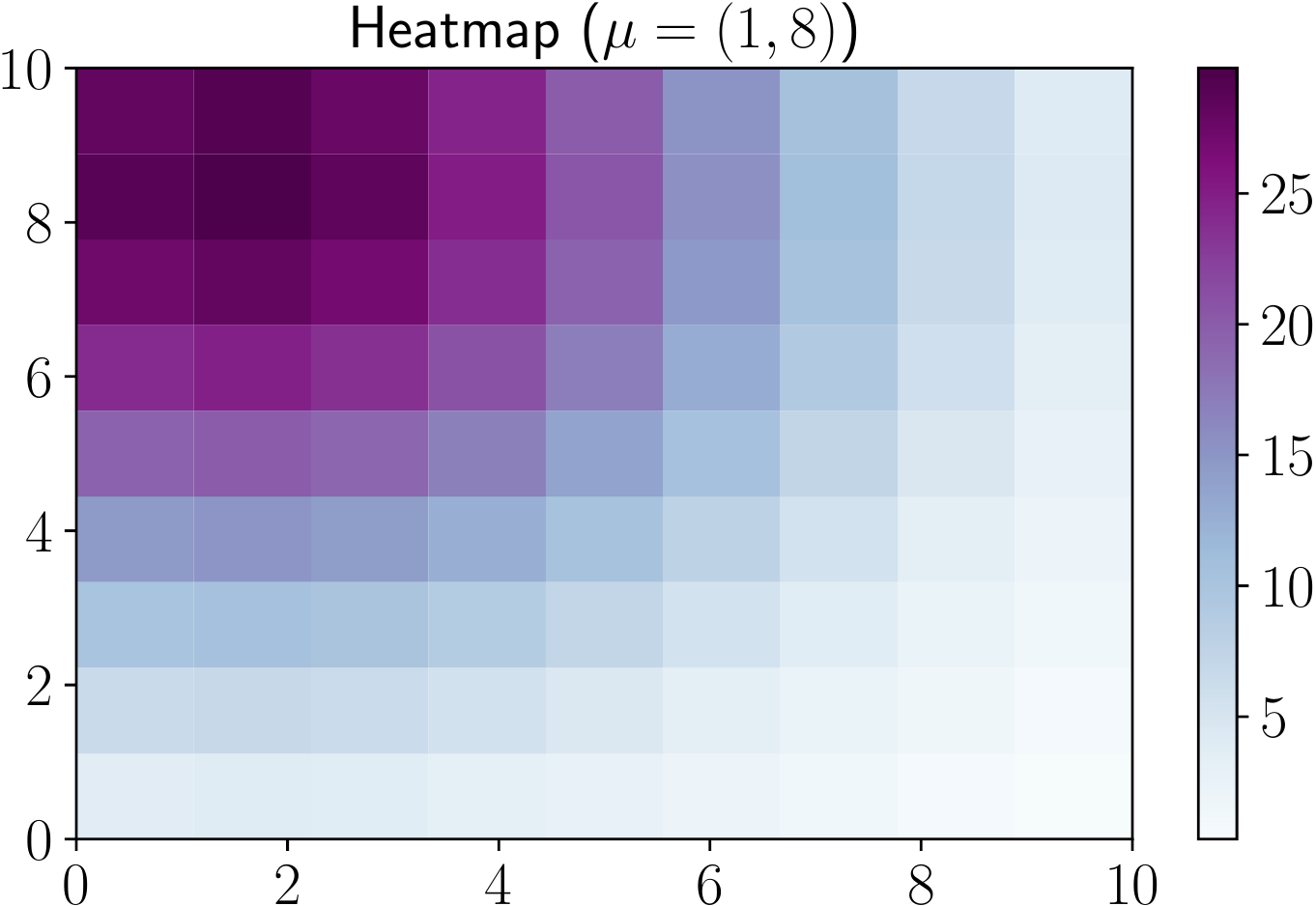}
            \caption{}
            \label{subfig:heatmap4}
        \end{subfigure}
        \caption{Order distribution changes. From (a) to (d), the order distribution of four consecutive timesteps in the $10 \times 10$ grid world is shown. The distance of $\mu_t$ and $\mu_{t+1}$ is 8 grids.}
        \label{fig:heatmap}
    \end{figure}

    In order to verify the feasibility of KL-divergence optimization, we adopt three cases correspond to different degree of order distribution changes. The degree of order distribution changes means the margin between two adjacent order distributions. In the particle order-dispatching environment, we generate orders with a given 2-dimensional Gaussian distribution $\mathcal{N}(\mu_t, \sigma_t)$ at each timestep. To quantify and explicitly compare the margin between order distributions at different timesteps, we can change $\mu_t$ at each timestep. The degree of order distribution changes is equivalent to the distance between adjacent order distributions. Farther distance means a higher degree, that is, the greater the degree of changes. In our experiment settings, the degree of changes from low to high orders correspond to a distance of 1, 2, 4 grids respectively,
    Figure~\ref{fig:heatmap} shows an example of order distribution changes, the distance between $\mu_t$ and $\mu_{t+1}$ is 8 grid. Since the destinations of orders are random, if we want to let vehicles serve more orders at the next timestep, we need to let the algorithm learns to pick suitable orders at current decision stage to assign the vehicles to suitable grids, so that we can ensure that there is a better ORR at next timestep. Intuitively, a long-term higher ORR corresponds to a higher long-term ADI. Table \ref{tb:toy_table} shows the performance at metrics of ORR and ADI at different degree of order distribution changes.
    
\subsubsection{Influence of $\lambda$}\label{sec:lambda}
    $\lambda$ plays an important role in our method, so it is necessary to investigate how it affects the performance at different degree of order distribution changes. In our experiments, the value of $\lambda$ ranges from 0.0 to 0.6 with stepping 0.05. When $\lambda=0.0$, it means our method is equivalent to IL. Figure~\ref{fig:lambda} shows curves at different degree of order distribution changes.

\subsubsection{Result Analysis}
    We train 300 episodes for all algorithms in the three cases which are related to different degrees of KL-divergence. We compare the three baselines from metrics of ADI and ORR. As shown in Table \ref{tb:toy_table}, it shows the average experimental results of 5 groups with different random seeds. The particle order-dispatching environment generates orders with random destinations, that is, the probabilities of long and short orders appear are equivalent in a grid. Although choosing long orders means higher ADI, the degree of order distribution changes we set requires the algorithm to choose more non-longest orders, thus ensuring both higher ORR and long-term ADI. In the three different degrees of order distribution changes, our KL-based outperforms all baselines on all metrics. It means our method can better counterweight the margin of order distribution. 
    
\begin{figure*}[htb]
        \centering
        \begin{subfigure}[b]{.3\textwidth}
            \centering
            \includegraphics[width=\textwidth]{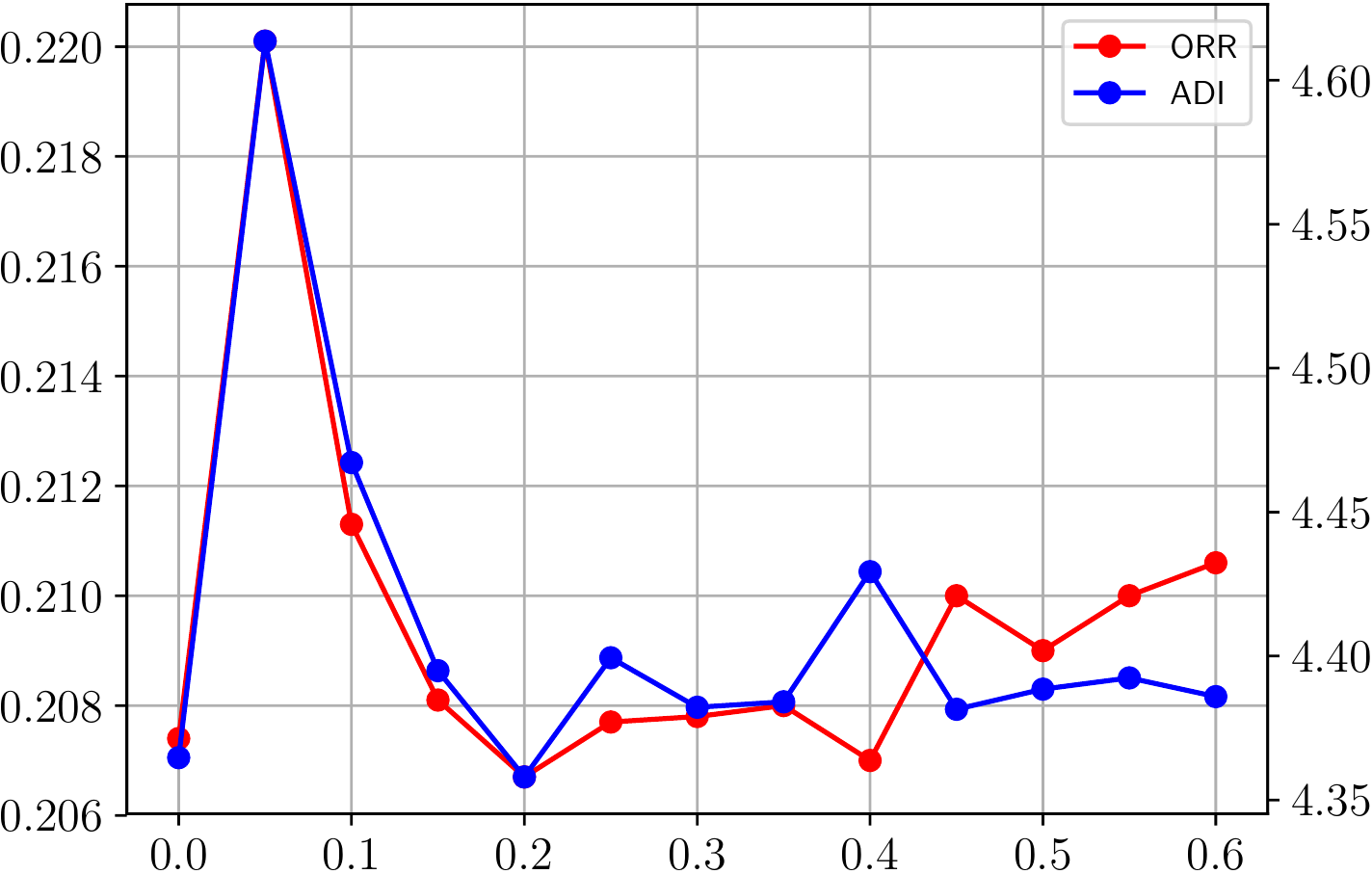}
            \caption{$\parallel \mu_t - \mu_{t+1}\parallel_2=1$}
            \label{fig:llambda}
        \end{subfigure}
        \hfill
        \begin{subfigure}[b]{.3\textwidth}
            \centering
            \includegraphics[width=\textwidth]{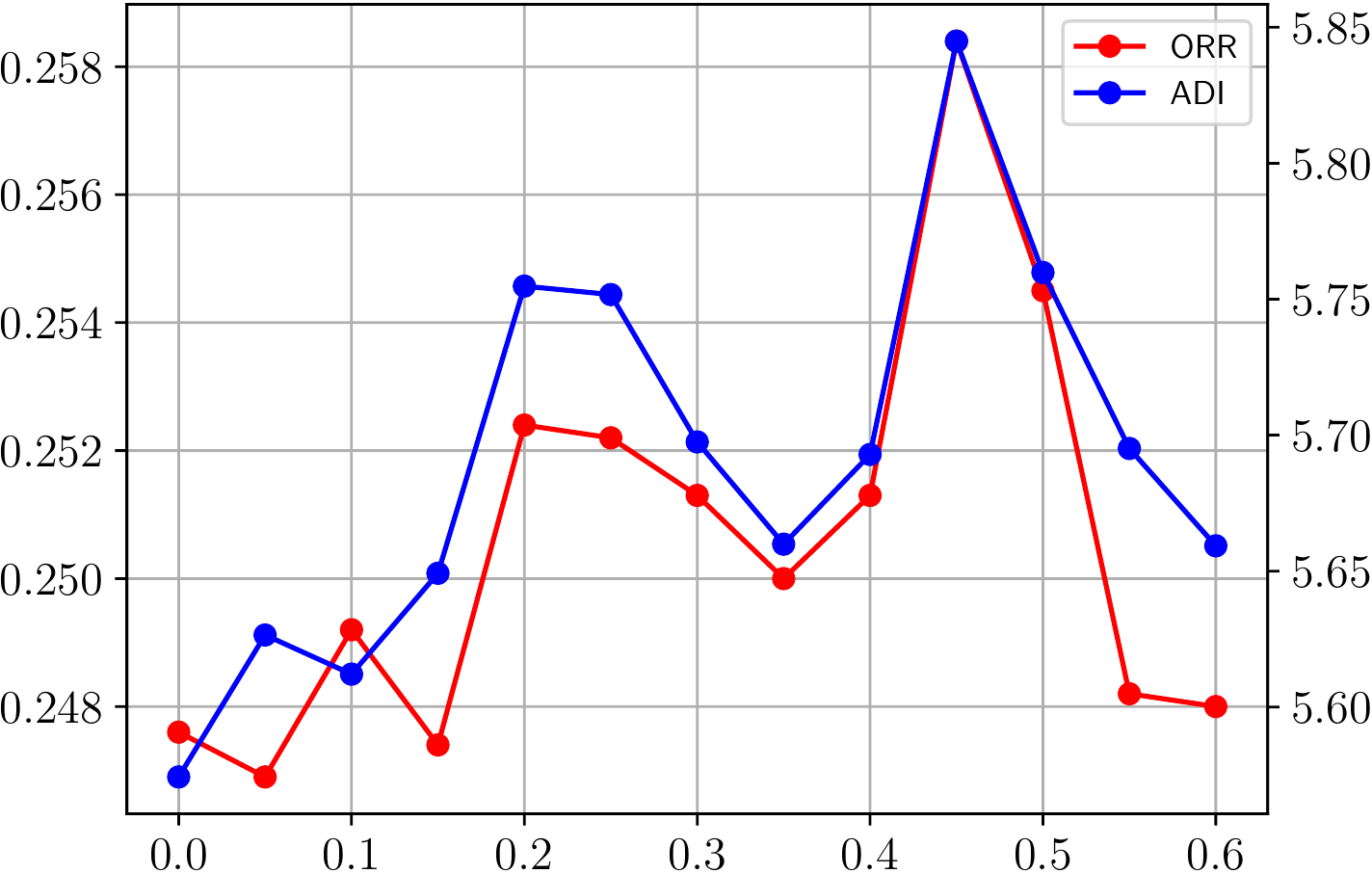}
            \caption{$\parallel \mu_t - \mu_{t+1}\parallel_2=2$}
            \label{fig:mlambda}
        \end{subfigure}
        \hfill
        \begin{subfigure}[b]{.3\textwidth}
            \centering
            \includegraphics[width=\textwidth]{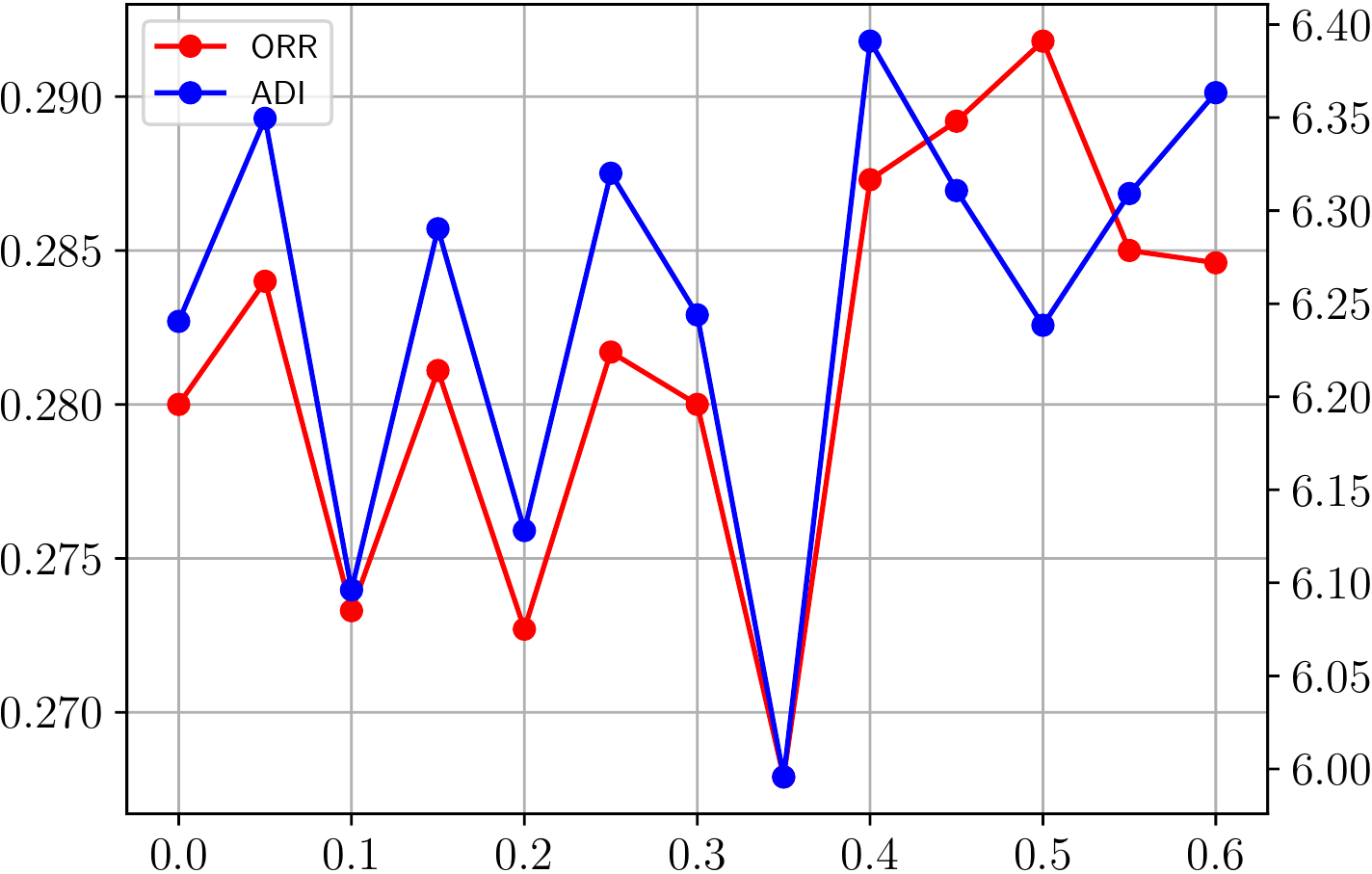}
            \caption{$\parallel \mu_t - \mu_{t+1}\parallel_2=4$}
            \label{fig:hlambda}
        \end{subfigure}
        \caption{ORR and ADI performance under different $\lambda$ settings. The horizontal axis represents different $\lambda$, the left and right vertical axis represent ORR and ADI respectively.}
        \label{fig:lambda}
    \end{figure*}

    Figure~\ref{fig:lambda} shows the learning curve of $\lambda$ at different degree of order distribution changes. In the three cases of order dispatching changes, our method achieves highest ORR at $\lambda=0.05$, $\lambda=0.45$, $\lambda=0.5$ respectively. The results in Figure~\ref{fig:lambda} also show that higher ORR often corresponds to higher ADI. In practice, if there is a low degree of order distribution changes, in order to achieve a higher ORR, the algorithm needs to pick shorter orders, so that in the future, agents in the closer regions can still serve more orders, namely, higher ORR. When it comes to a high degree of order distribution changes, the dispatching algorithm needs to perform more greedy to achieve a higher ORR, that is, it prefers to pick longer orders. Also, the result of Figure~\ref{fig:hlambda} shows our algorithm achieves better performance on ORR and ADI than IL in the case of a high degree of order distribution changes. Therefore, combined with the results of Table.~\ref{tb:toy_table} and Figure~\ref{fig:lambda}, our method can flexibly choose long or short orders.

\subsection{Real World Data Experiments}
\subsubsection{Dispatching Simulator}
    Since our model is implemented on the setting of dividing the city into many order dispatching regions, so we conduct experiments on an open source grid-based environment simulator provided by Didi Chuxing \cite{lin2018efficient}. The simulator divides the city into $N$ hexagonal grids which depends on the size of the city. At each time $t$, the simulator provides a set of idle vehicles and a set of available orders. Each order is featured with its origin, destination, and duration, and vehicles in the same region share the same state. The travel distance between neighboring regions is approximately 2.2km and the time interval is 10min.

\subsubsection{Data Description}
    The real-world datasets provided by Didi Chuxing include order and trajectories of vehicles information of three cities in one month. The order information includes price, origin, destination, and duration. The trajectories contain the positions (latitude and longitude) and status (on-line, off-line, on-service). We divide the three cities into 182, 126, 112 hexagonal grids respectively.

    \paragraph{Result Analysis} We compare our model with three baselines after 300 episodes training. As shown in Table \ref{tb:real_world_exp}, it lists the average results of 5 groups experiments with different random seeds. The real datasets contain more changes in the order distribution. From the results, our method can still better discover the changes of order distribution and improve the ORR and ADI via order-vehicle distribution matching.
    
\section{Deployment}
\begin{figure}[h]
        \centering
        \includegraphics[width=\columnwidth]{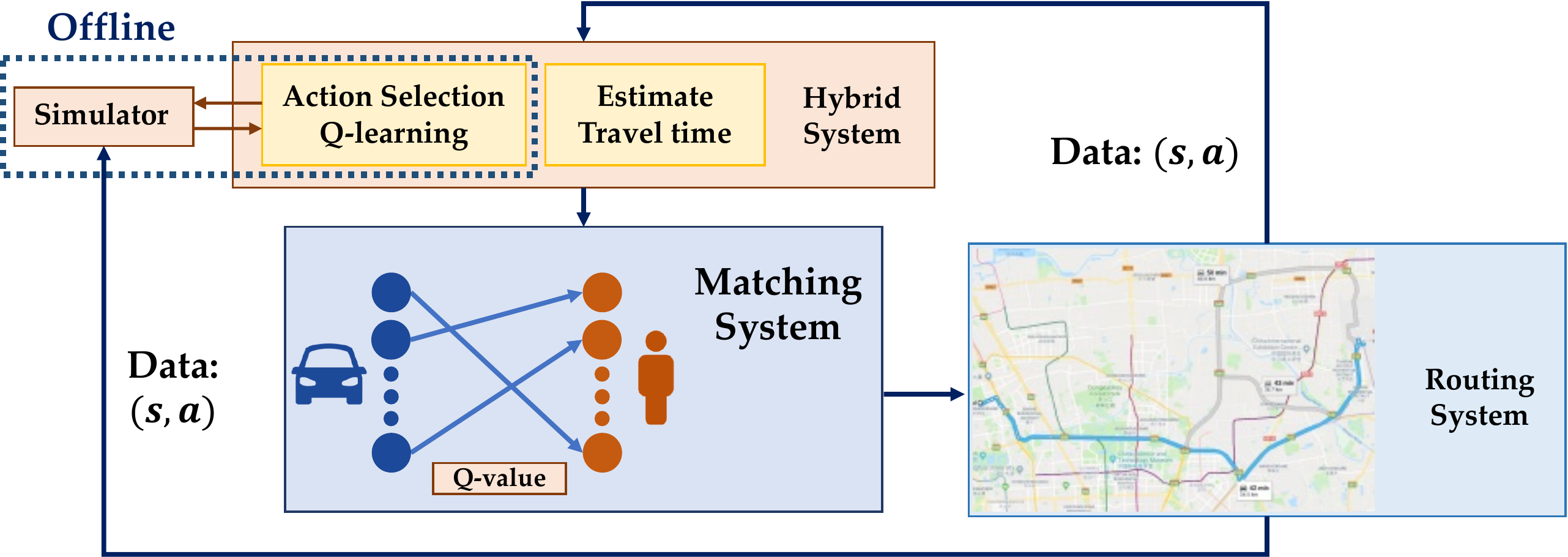}
        \caption{Illustration of deployment. The hybrid system consists of two modules, namely, Action Selection Q-learning (ASQ) and Estimate Travel time modules. The ASQ will interact with simulator periodically, and it will be trained offline in the simulator. Matching System accepts value estimation and outputs $\langle \text{vehicle}, \text{order} \rangle$ matches to Routing System.}
        \label{fig:implementation}
    \end{figure}

    Taking both model setting in this paper and online platforms of Didi Chuxing, we design a hybrid system and incorporate with other components including routing planning technique \cite{tong2018unified} and estimating time of arrival (ETA) \cite{wang2018learning} as illustrated in Figure~\ref{fig:implementation}.\par

    As aforementioned mentioned in Section~\ref{sec:method}, there are several assumptions prevent this model from deploying in real-world settings: (i) vehicles in the same grid share the same setting, and this isomorphic setting ignores the intra-grid information; (ii) this paper adopts the grid-world map to simplify the real-world environment which replace coordinate position information with grid information. To address these issues, we adapt estimate travel time techniques proposed in and incorporate with our action selection Q-learning mentioned in Section~\ref{subsec:qlearning}. For example, the duration time of each order in our model is regarded as one of the already known order features. However, in the real-world scenario, each order's travel time obtained via the ETA model is dynamic and depends on current traffic and route conditions. Since coordinate position information is taken into consideration in the ETA model, this hybrid system is able to deal with the assumption (ii) and feasible to be deployed in real-world.
    
    We extend the Matching System and the Routing System after obtaining $Q$-value via the hybrid system as illustrated in Figure.~\ref{fig:implementation}. Specifically, in each time slot, the goal of the real-time order dispatch algorithm is to determine the best matching between vehicles and orders (see Figure~\ref{fig:order_dispatching}) in the matching system and plan a routine for drivers to serve the orders. Formally, the principle of Matching System can be formulated as:
    
    \begin{equation}
        \label{equ:km}
        \text{argmax}_{a_{ij}} \sum_{i=0}^{m} \sum_{j=0}^{n} Q(i,j)a_{ij}~,
    \end{equation}

    \begin{equation}
        \label{equ:cons}
        \begin{aligned}
        s.t. 
        & \sum_{i=0}^m a_{ij}=1,\ j=1,2,3...,n \\
        & \sum_{j=0}^{n} a_{ij}=1, i=1,2,3...,m
        \end{aligned}
    \end{equation}
    where
    $$
    a_{ij} = 
    \begin{cases}
    1,  & \text{if order $j$ is assigned to driver $i$} \\
    0, & \text{if order $i$ is not assigned to driver $i$}
    \end{cases}
    $$
    where $i \in [1,2,...,m]$ and $j \in [1,2,...,n]$ present all idle drivers and available orders at each time step respectively. $Q(i,j)$ is the output from hybrid system and represents the action-value function driver $i$ performing an action of serving order $j$. Note that constraints in Eq.~(\ref{equ:cons}) guarantee that each driver will select one available real orders or doing nothing while each order will be assigned to one driver or stay unserved at each time step. 
    
    This Matching System used in \citet{xu2018large} and \citet{wang2018deep} is implemented using Kuhn-Munkres (KM) algorithm \cite{munkres1957algorithms}. In detail, they formulated Eq.~(\ref{equ:km}) as a bipartite graph matching problem where drivers and orders are presented as two set of nodes. Then, each edge between order $i$ and driver $j$ is valued with $Q(i,j)$, and the best matches will be fined using KM algorithm. Different from them, since we implemented our method based on assumption (i), that is, there is no difference in the drivers in a same grid. So the KM algorithm will degenerate into a sorting algorithm here. We just need to select the top $m$ orders with the highest $Q(i,j)$ values.

    Once the matching pairs of orders and vehicles has been selected from the matching system, we then deliver these pairs with coordinate information to the routing system. The routing system equipped with route planning techniques \cite{tong2018unified} allows drivers to serve the order. This process will give feedback, i.e. reward to the hybrid system and help the whole system training to achieve better performance.

\section{Conclusions} 
    In this paper, we proposed a multi-agent reinforcement learning method for order-dispatching via matching the distribution of orders and vehicles. Results on the three cases in the simulated order-dispatching environment have demonstrated that our proposed method achieves both higher ADI and ORR than the three baselines, including one independent MARL method, one planning algorithm, and one rule-based algorithm, in various traffic environments. The experiments on real-world datasets also show that our model can obtain higher ADI and ORR. Furthermore, our proposed method is a centralized training method and can be executed decentralized. In addition, we designed the deployment system of the model with the support of the existing platform of Didi Chuxing. In future work, we plan to deploy the model to do online tests through the designed deployment system.

\section*{Acknowledgments}
    We thank the support of National Natural Science Foundation of China
    (61702327, 61772333, 61632017).

%\clearpage
\bibliographystyle{ACM-Reference-Format}
\balance
\bibliography{references}

\end{document}